\documentclass[twocolumn]{aa}
\usepackage{astrobib}
\usepackage{journals}
\usepackage{graphicx}
\usepackage{txfonts}
\begin{document}
   \title{Water masers in the Local Group of galaxies}

   \subtitle{}

   \author{A. Brunthaler
          \inst{1,2}
          \and
          C. Henkel\inst{1}
          \and
          W.J.G. de Blok\inst{3}
          \and
          M.J. Reid\inst{4}
	  \and 
	  L.J. Greenhill\inst{4}
	  \and
          H. Falcke\inst{5,6}
          }

   \offprints{brunthal@mpifr-bonn.mpg.de}

   \institute{Max-Planck-Institut f\"ur Radioastronomie, Auf dem H\"ugel 69,
              53121 Bonn, Germany
              \and
              Joint Institute for VLBI in Europe, Postbus 2, 7990 AA
              Dwingeloo, The Netherlands
              \and
              Research School of Astronomy and Astrophysics, Australian National University,
              Mount Stromlo Observatory, Cotter Road, Weston Creek, ACT 2611, Australia
              \and
	      Harvard-Smithsonian Center for Astrophysics, 60 Garden Street,
              Cambridge, MA 02138, USA
              \and
              ASTRON, Postbus 2, 7990 AA Dwingeloo, the Netherlands
	      \and 
	       Department of Astrophysics, Radboud Universiteit
               Nijmegen, Postbus 9010, 6500 GL Nijmegen, The Netherlands}

   \date{Received; accepted}

   \abstract{
We compare the number of detected 22 GHz H$_2$O masers in the Local Group 
galaxies M31, M33, NGC\,6822, IC\,10, IC\,1613, DDO\,187, GR8, NGC\,185, and 
the Magellanic Clouds with the water maser population of the Milky Way. 
To accomplish this we searched for water maser emission in the two Local Group 
galaxies M33 and NGC\,6822 using the Very Large Array (VLA) and incorporated 
results from previous studies. We observed 62 H{\sc ii} regions in 
M33 and 36 regions with H$\alpha$ emission in NGC\,6822. Detection limits are 
0.0015 and 0.0008 L$_\odot$ for M33 and NGC\,6822, respectively (corresponding 
to 47 and 50 mJy in three channels with 0.7 km~s$^{-1}$ width). M33 hosts three
water masers above our detection limit, while in NGC\,6822 no maser source was 
detected. We find that the water maser detection rates in the Local Group 
galaxies M31, M33, NGC\,6822, IC\,1613, DDO\,187, GR8, NGC\,185, and the 
Magellanic Clouds are consistent with 
expectations from the Galactic water masers if one considers the different 
star formation rates of the galaxies. However, the galaxy IC\,10 exhibits an
overabundance of masers, which may result from a compact 
central starburst.

   }

   \maketitle
%

\section{Introduction}

Galactic water maser emission, discovered by \citeN{CheungRankTownes1969}, is 
mainly associated with star forming regions and late type stars. In star 
forming regions the masers are found in the vicinity of compact H{\sc ii} 
regions (\citeNP{GenzelDownes1977}). Observations of these maser sources 
trace extremely dense ($>10^7$cm$^{-3}$) and warm ($>400$ K) molecular 
gas and yield important information about physical and dynamical properties 
of their environments. They can also be used to measure accurate annual 
parallaxes and proper motions of star forming regions using Very Long Baseline 
Interferometry (VLBI) (e.g. \citeNP{HachisukaBrunthalerReid2006}). While more 
than one thousand water masers are known in the Milky Way 
\cite{ValdettaroPallaBrand2001}, only a small number are known in other Local 
Group galaxies.

The first extragalactic water maser was discovered by 
\citeN{ChurchwellWitzelHuchtmeier1977} in IC\,133, a star forming region 
in the Local Group galaxy M33. Further water masers in  Local Group galaxies 
were found towards the Magellanic Clouds (e.g.~\citeNP{ScaliseBraz1981}) and 
IC\,10 (e.g. \citeNP{BeckerHenkelWilson1993}). Searches toward M31 
\cite{GreenhillHenkelBecker1995,ImaiIshiharaKameya2001} and other Local Group
galaxies \cite{HuchtmeierRichterWitzel1980,GreenhillMoranReid1990} have not 
yet yielded a detection. Beyond the Local Group, water masers are found mainly 
toward active galactic nuclei (see \citeNP{ZhangHenkelKadler2006}, and 
\citeNP{Kondratko2006} for a list, and 
http://www.cfa.harvard.edu/wmcp/surveys/survey.html for an updated list).

Proper motion measurements and accurate distances of Local Group galaxies are 
important for our understanding of the dynamics and evolution of the Local 
Group. With the NRAO\footnote{The National Radio Astronomy Observatory is 
operated by Associated Universities, Inc., under a cooperative agreement with 
the National Science Foundation.} Very Long Baseline Array (VLBA) 
\citeN{BrunthalerReidFalcke2005} 
measured the proper motions of two regions of water maser activity on 
opposite sides of M33. The comparison of the relative proper motion between 
the two regions and the expected motion from the known rotation curve and 
inclination of M33 led to a determination of a rotational parallax 
(730 $\pm$ 168 kiloparsecs) of this galaxy. This distance is consistent 
with recent Cepheid and tip of the red giant branch estimates 
(\citeNP{LeeKimSarajedini2002}; \citeNP{McConnachieIrwinFerguson2005}) and 
earlier distance estimates using the internal motions of water masers in 
IC\,133 \cite{GreenhillMoranReid1993,ArgonGreenhillMoran2004}. Since 
the proper motion measurements were made relative to a distant extragalactic 
background source, the proper motion of M33 itself could also be determined.  
This measured proper motion of M33 is a first important step towards a 
kinematical model of the Local Group and was used to constrain the proper 
motion of the Andromeda Galaxy M31 \cite{LoebReidBrunthaler2005}.

The interpretation of the proper motions of water masers in Local Group
galaxies may be limited by the unknown three dimensional peculiar motions 
of the star forming regions. In the Milky Way, peculiar motions of 
star forming regions can be 20 km~s$^{-1}$ as seen in W3(OH) 
(\citeNP{XuReidZheng2006}). Hence it is important to use multiple regions of 
maser activity for proper motion studies to constrain the systematic effect of
peculiar motions. This lead us to search for water vapor masers towards 62 
H{\sc ii} regions in M33 with the NRAO Very Large Array (VLA). 

A detection of maser sources in other Local Group galaxies would allow us to 
extend such studies considerably. Since the strongest maser emission from high-mass star froming regions in our Galaxy is typically 100 times stronger than emission from late-type stars, we targeted NGC\,6822, a northern Local Group member with prominent star formation
activity.

\section{Observations}

\subsection{M33}

\begin{figure}
\resizebox{\hsize}{!}{\includegraphics[]{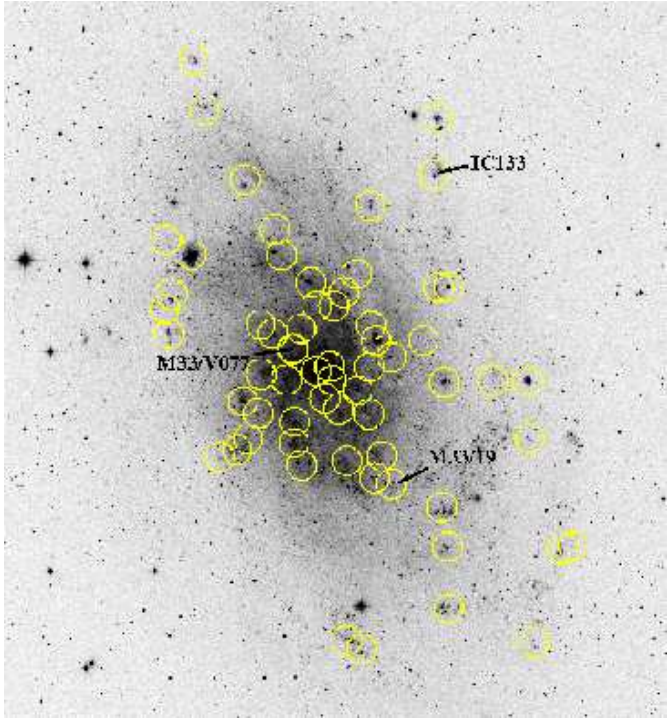}}
\caption{DDS image of M33 with the positions of the 62 observed H{\sc ii} 
regions. The diameter of the circles is 2 arcminutes, corresponding to the primary beam of 
the VLA antennas.}
\label{pos_m33}
\end{figure}

We selected H{\sc ii} regions which also have H{\sc ii} velocity measurements 
(\citeNP{ZaritskyElstonHill1989}) and are also identified through their radio 
continuum emission by \citeN{ViallefondGossvanderHulst1986} with flux densities
above 0.5 mJy at 1.4 GHz. We also added NGC\,588 and six H{\sc ii} regions 
from \citeN{ViallefondGossvanderHulst1986} that are either bright at radio 
frequencies ($>$1.8 mJy) or are located in the outer parts of M33 (the last 
six sources in Table~\ref{source1}). The two maser sources in M33 from our 
VLBA observations \cite{BrunthalerReidFalcke2005}, M33/19 and IC\,133 , were  
observed to ensure correct telescope and receiver operation and to investigate
time variabilty. The 62 observed fields are listed in Tables~\ref{source1} and 
\ref{source2} and are shown in Fig.~\ref{pos_m33}. Here the source name 
M33/V001 stands for source 001 from the source list in 
\citeN{ViallefondGossvanderHulst1986}.

The observations were made with the VLA in A-configuration on 15 September 
and 18 November 2004 and lasted 7.5 and 6 hours, respectively. The source 
3C\,48 was the primary flux density calibrator. We used fast switching, 
spending 120 seconds on the target and 60 seconds on a reference source 
(3C\,48) for a total of 12 minutes for each source in the first observation 
and 9 minutes in the second observation. The integration time was 3 seconds. 
The observations were made with a total bandwidth of 6.25 MHz and 128 
spectral channels corresponding to a channel spacing of $\sim$0.7 km s$^{-1}$. 
The band was centered at the velocity of the H{\sc ii} gas for sources 
listed in \citeN{ZaritskyElstonHill1989}. For sources with no velocity 
information, we calculated the expected radial velocity using the rotation 
model of M33 by \citeN{CorbelliSchneider1997}. The uncertainty of the model is
less than 15 km~s$^{-1}$ (see discussion in \citeNP{BrunthalerReidFalcke2005}) 
and much smaller than the covered velocity range of 84 km~s$^{-1}$.

\subsection{NGC\,6822}

\begin{figure}
\resizebox{\hsize}{!}{\includegraphics[]{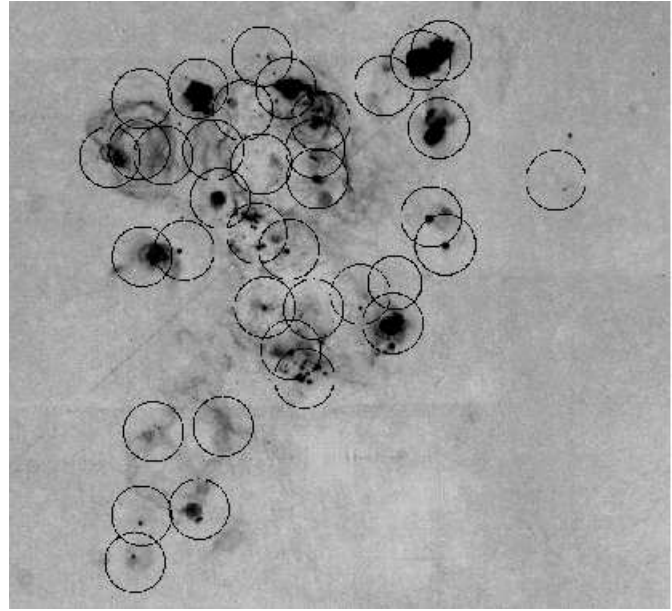}}
\caption{The H$\alpha$ distribution in the central part of NGC\,6822 (taken 
from \protect\citeNP{deBlokWalter2006}) with the positions of the observed 
regions. The diameter of the circles is 2 arcminutes, the primary beam of 
the VLA antennas.}
\label{pos_ngc6822}
\end{figure}

NGC\,6822 was observed with the VLA in A-configuration on 13 March (2 hours), 
25 March (2.5 hours), 31 March (2.5 hours), 10 April (2 hours), and 17 April 
2006 (2.5 hours). We observed a total of 36 regions with H\,$\alpha$ emission 
(Fig.~\ref{pos_ngc6822} and Table~\ref{ngc6822_source}). 3C\,48 was also the 
primary flux density calibrator. The observations were similar to the 
observations of M33 with fast switching between the target (120 seconds) and 
the reference source 1939-154 (60 seconds) for a total of 12 minutes.
We used the H\,I data from \citeN{deBlokWalter2006} to estimate the radial 
velocities of the gas at the different pointings.

\begin{figure}
\resizebox{\hsize}{!}{\includegraphics[bbllx=9.0cm,bburx=19.5cm,bblly=2.2cm,bbury=17.0cm,clip=,angle=-90]{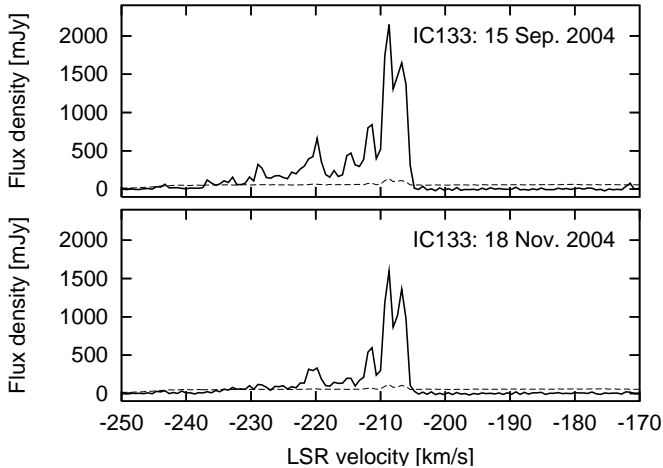}}
\caption{Spectra of IC\,133 Main for the observations on 15 September 2004 
(upper panel) and 18 November 2004 (lower panel). The dashed lines 
are the 5$\sigma$ detection limits in a single channel.
}
\label{ic133-spec}
\end{figure}

\begin{figure}
\resizebox{\hsize}{!}{\includegraphics[bbllx=8.8cm,bburx=19.5cm,bblly=2.2cm,bbury=17.0cm,clip=,angle=-90]{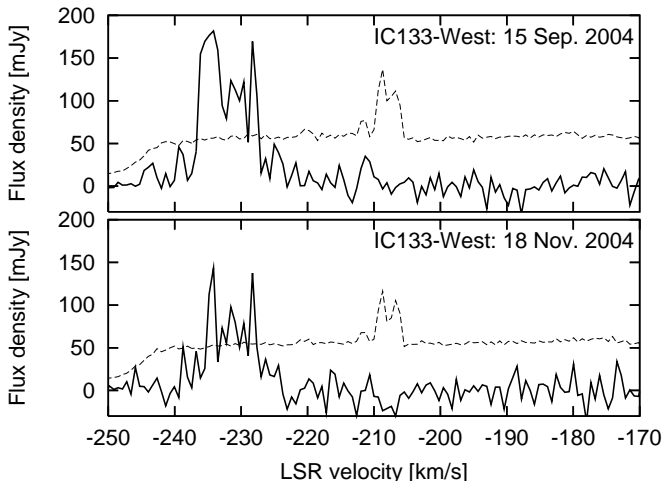}}
\caption{Spectra of IC\,133-West for the observations on 15 September 2004 
(upper panel) and 18 November 2004 (lower panel). The dashed lines 
are the 5$\sigma$ detection limits in a single channel. 
}
\label{ic133w-spec}
\end{figure}

\section{Data Reduction and Results}

The data for both galaxies were edited and calibrated with standard techniques
in the Astronomical Image Processing System (AIPS). We used a source model 
generated by the AIPS task CALRD for the phase calibration on 3C\,48 and a 
point source model for 1939-154. The primary beam of the 
VLA antennas at a frequency of 22 GHz has a full-width at half-maximum of 
$\approx$ 2 arcminutes. We mapped all frequency channels 
for each telescope pointing to half-power response of the primary beam 
with 4096$\times$4096 pixels of 0.03". The resulting data cubes were 
then searched for emission. We considered emission as real only if 
detected in at least three contiguous frequencies channels with a signal to 
noise ratio of $>5 \sigma$ or in a single channel at $>7 \sigma$.

\subsection{M33}

\begin{figure}
\resizebox{\hsize}{!}{\includegraphics[bbllx=13.0cm,bburx=19.5cm,bblly=2.0cm,bbury=17.0cm,clip=,angle=-90]{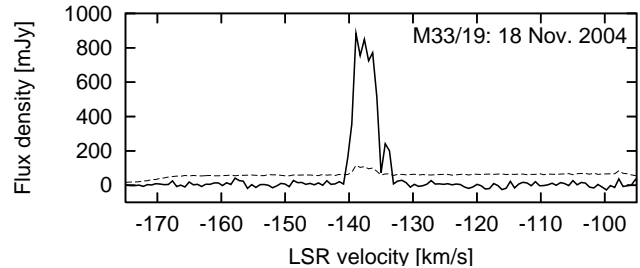}}
\caption{Spectrum of M33/19 for the observation on 18 November 2004. The 
dashed line is the 5$\sigma$ detection limit in a single channel.
}
\label{m33_19-spec}
\end{figure}

\begin{figure}
\resizebox{\hsize}{!}{\includegraphics[bbllx=9.0cm,bburx=19.5cm,bblly=2.0cm,bbury=17.0cm,clip=,angle=-90]{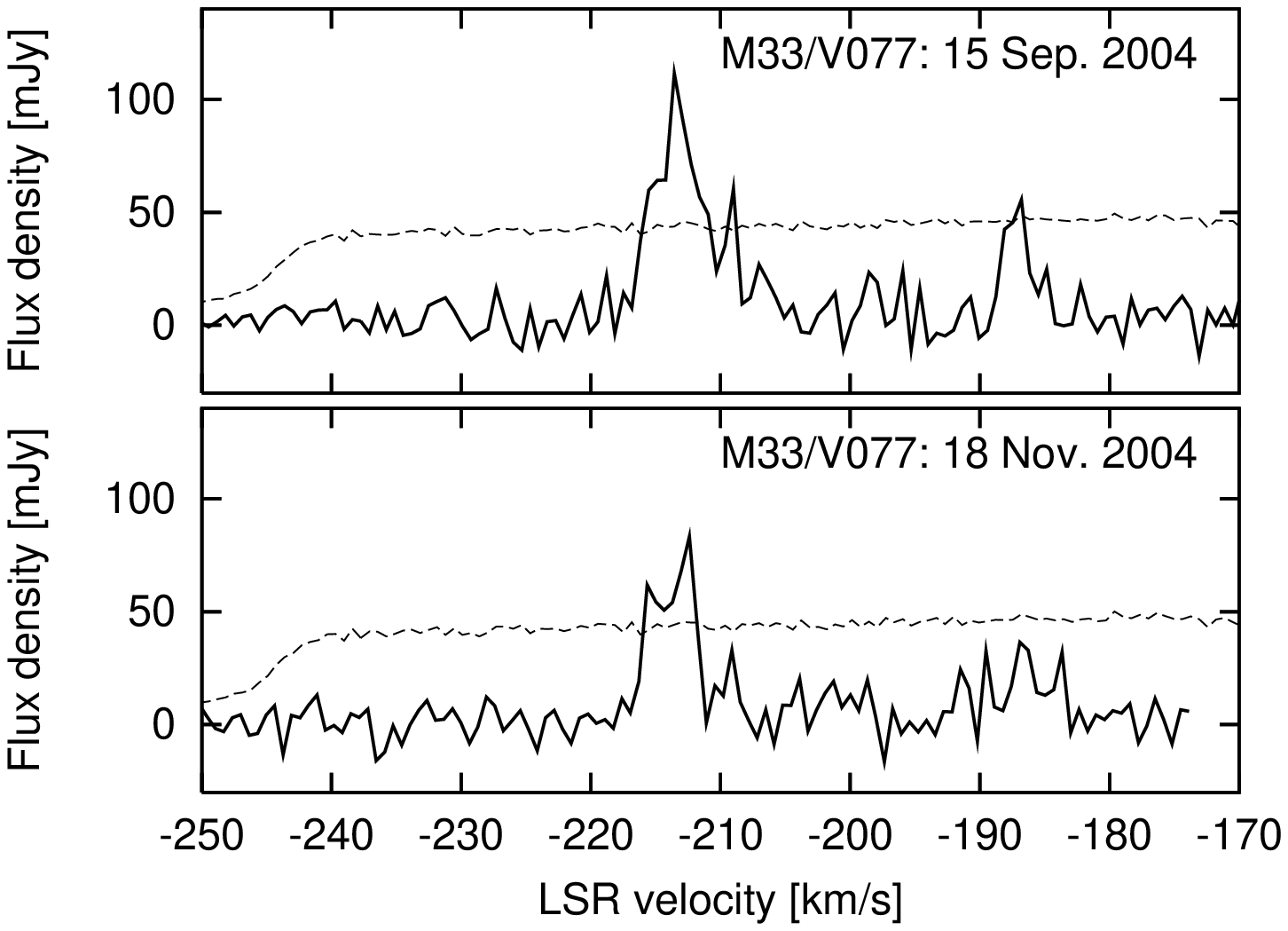}}
\caption{Spectra of M33/V077 for the observations on 15 September 2004 (upper 
panel) and 18 November 2004 (lower panel). The dashed lines are 
the 5$\sigma$ detection limits in a single channel.}
\label{m33_v77-spec}
\end{figure}

We observed maser emission in two well known H{\sc ii} regions, namely M33/V028
(IC\,133), and M33/V033 (M33/19). These two maser sources have been studied in detail previously (see 
\citeNP{GreenhillMoranReid1993,ArgonGreenhillMoran2004,BrunthalerReidFalcke2005}). Other tentative water maser detections were reported in 
\citeN{HuchtmeierRichterWitzel1980} and \citeN{HuchtmeierEckartZensus1988}, 
of which we only confirm emission from M33/V077 (listed as M33/50 in 
\citeNP{HuchtmeierEckartZensus1988}). For this source we obtain a position of:

\begin{equation}
\alpha_{J2000}= \mathrm{01^h~34^m~00.23^s}\\
\delta_{J2000}= \mathrm{+30^\circ~40'~47.3"}
\end{equation}
with an uncertainty of 0.1 arcseconds. 

For all other regions we derived 
$5\sigma$ upper limits between 37 and 65 mJy per 0.7 km~s$^{-1}$ 
(see Tables \ref{source1} and \ref{source2}) depending on the observing 
time and amount of flagged data, with an average value of 47 mJy.  We 
did not detect a maser in M33/16 that was reported  based on Effelsberg 
single-dish data in \citeN{HuchtmeierEckartZensus1988}. M33/16 is 
located less than 30 arcseconds from the known maser in M33/19, so that an 
interferometer is needed to clearly discriminate between the two sources. 
The maser in M33/2 at $-162$ km~s$^{-1}$ 
\cite{HuchtmeierWitzelKuehr1978,HuchtmeierRichterWitzel1980} can not be 
confirmed. 

In IC\,133 the emission is found in two spatially distinct regions separated 
by $\sim$0.3 arcseconds, IC\,133-Main and IC\,133-West.
The spectra of all detected sources are shown in Figs.~\ref{ic133-spec},
~\ref{ic133w-spec}, \ref{m33_19-spec}, and \ref{m33_v77-spec}. IC\,133 and M33/V077 were observed 
in both observing periods, while M33/19 was observed only during the second 
period. IC\,133 and M33/V077 show variability between the two observations. 
The peak flux densities of both sources dropped by $\sim$ 25\%. For individual 
channels we find variations between 0 and $-70$ \% in IC\,133 and between +70 
and $-50$\% in M33/V077. The large differences between individual channels show 
clearly that a major part of the variations are intrinsic to the maser 
sources and not caused by instrumental effects. Similar variations are  
common in Galactic water maser sources (e.g. 
\citeNP{BrandCesaroniComoretto2005}) or in IC\,10, where even intraday 
variations were found (\citeNP{ArgonGreenhillMoran1994}) although interstellar
scintillation can not be ruled out in IC\,10.

\begin{figure}
\resizebox{\hsize}{!}{\includegraphics[bbllx=8.0cm,bburx=19.5cm,bblly=1.9cm,bbury=17.0cm,clip=,angle=-90]{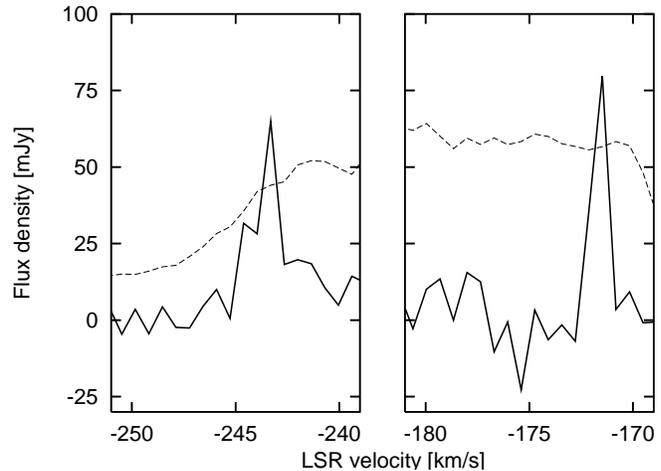}}
\caption{Blow-up of the weak emission around -243 km~s$^{-1}$ (left) and 
-171 km~s$^{-1}$ (right) from the observations of IC\,133 on 15 September 2004.
The dashed lines are the 5$\sigma$ detection limits in a single 
channel.}
\label{ic133-spec-hv}
\end{figure}

The spectrum of IC\,133-Main has a peak near $-210$ km~s$^{-1}$ and a 
one-sided broad wing extending down to $\sim$240 km~s$^{-1}$ wich is similar 
to published spectra in the literature (see e.g. 
\citeNP{HuchtmeierEckartZensus1988} for several spectra between 1976 and 1985).
The peak flux densities vary between 1 and 2 Jy and new features appear in the 
broad wing towards lower radial velocities on timescales of months to years. 
However, in the first epoch we detect weak emission with flux densities of 
65 and 79 mJy (7.4 and 7.2 $\sigma$)at $-243$ km~s$^{-1}$
and $-171$ km~s$^{-1}$, respectively (see Fig~\ref{ic133-spec-hv}). This weak 
emission is almost symmetrical to the strongest peak at $-208$ 
km~s$^{-1}$. The detection of emission at $-171$ km~s$^{-1}$ is to our 
knowledge the first time that emission at significantly higher radial velocities
than the strong peak near $-210$ km~s$^{-1}$ has been seen. 

\begin{figure}
\resizebox{\hsize}{!}{\includegraphics[bbllx=9.0cm,bburx=19.5cm,bblly=2.0cm,bbury=17.0cm,clip=,angle=-90]{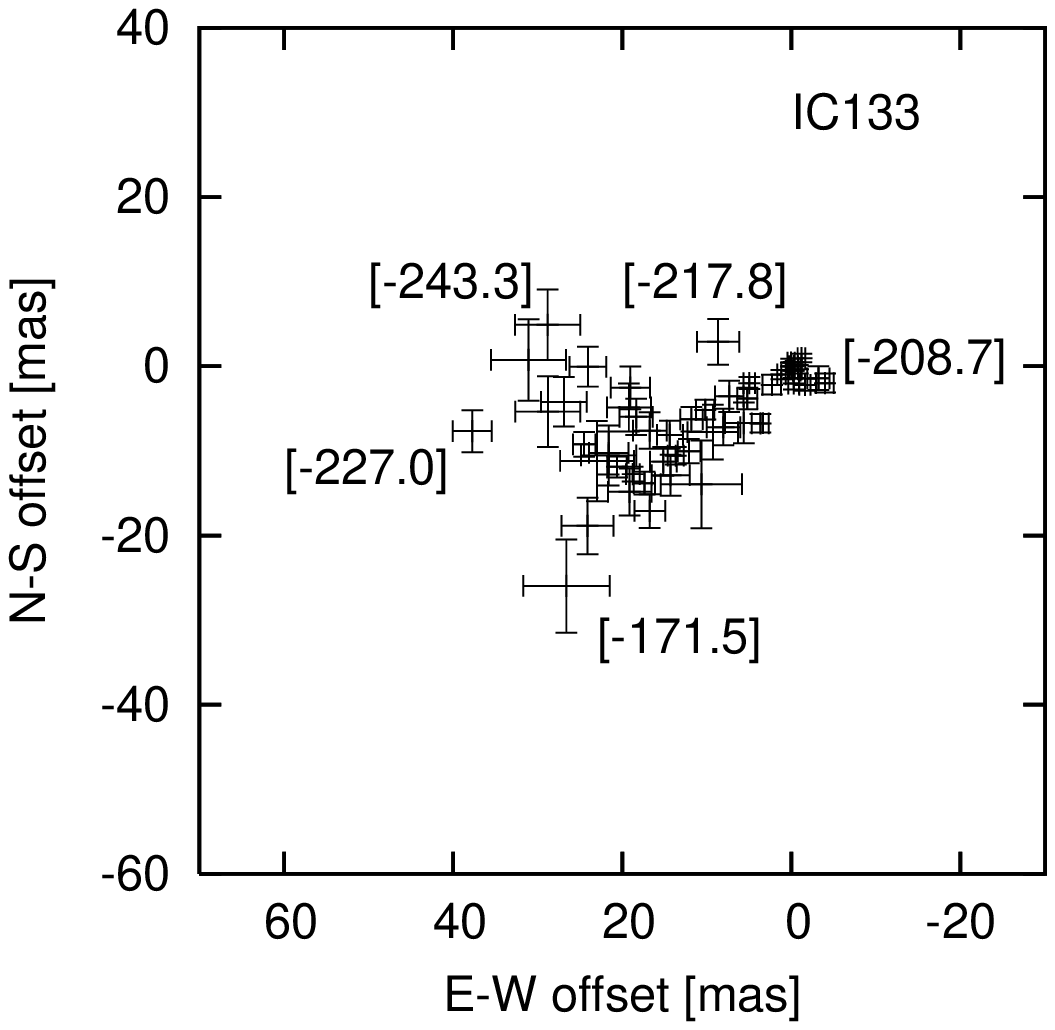}}
\resizebox{\hsize}{!}{\includegraphics[bbllx=9.0cm,bburx=19.5cm,bblly=2.0cm,bbury=17.0cm,clip=,angle=-90]{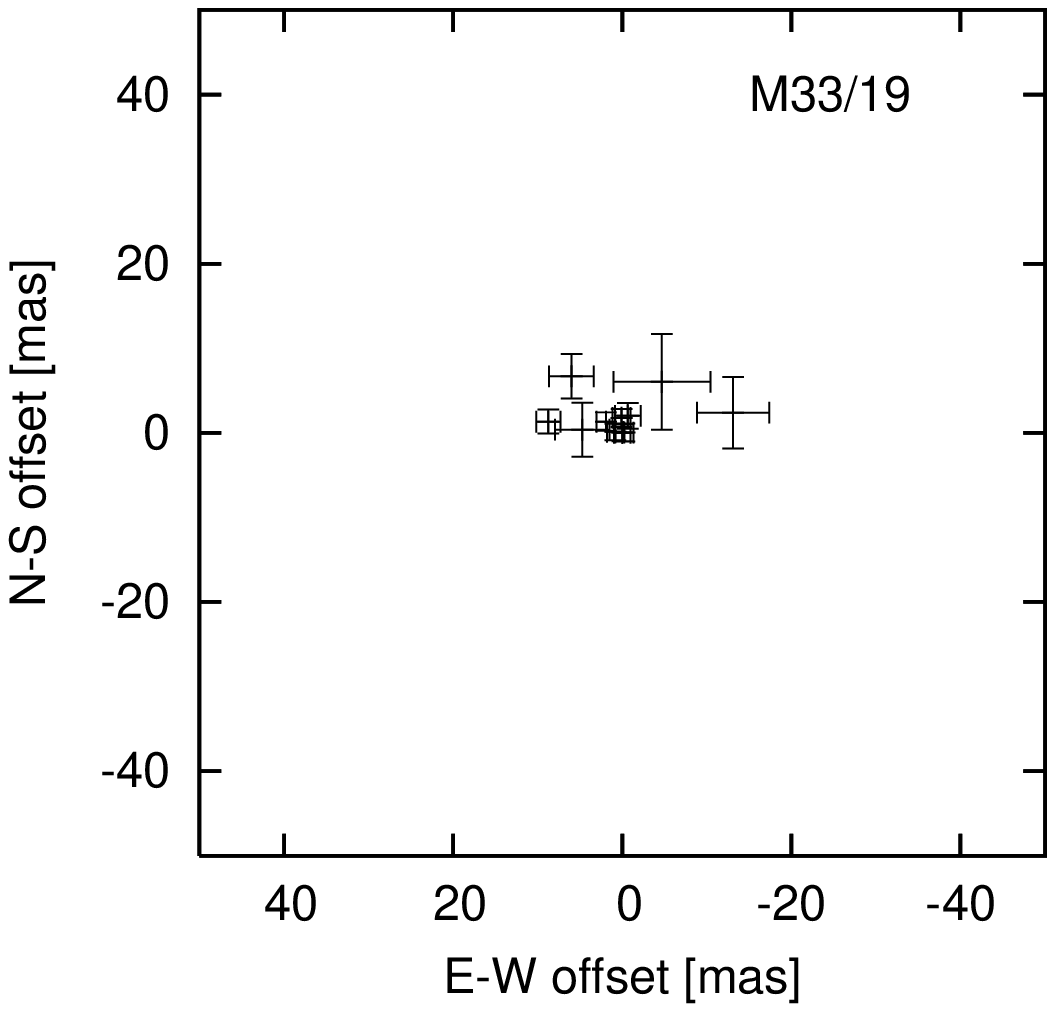}}
\resizebox{\hsize}{!}{\includegraphics[bbllx=9.0cm,bburx=19.5cm,bblly=2.0cm,bbury=17.0cm,clip=,angle=-90]{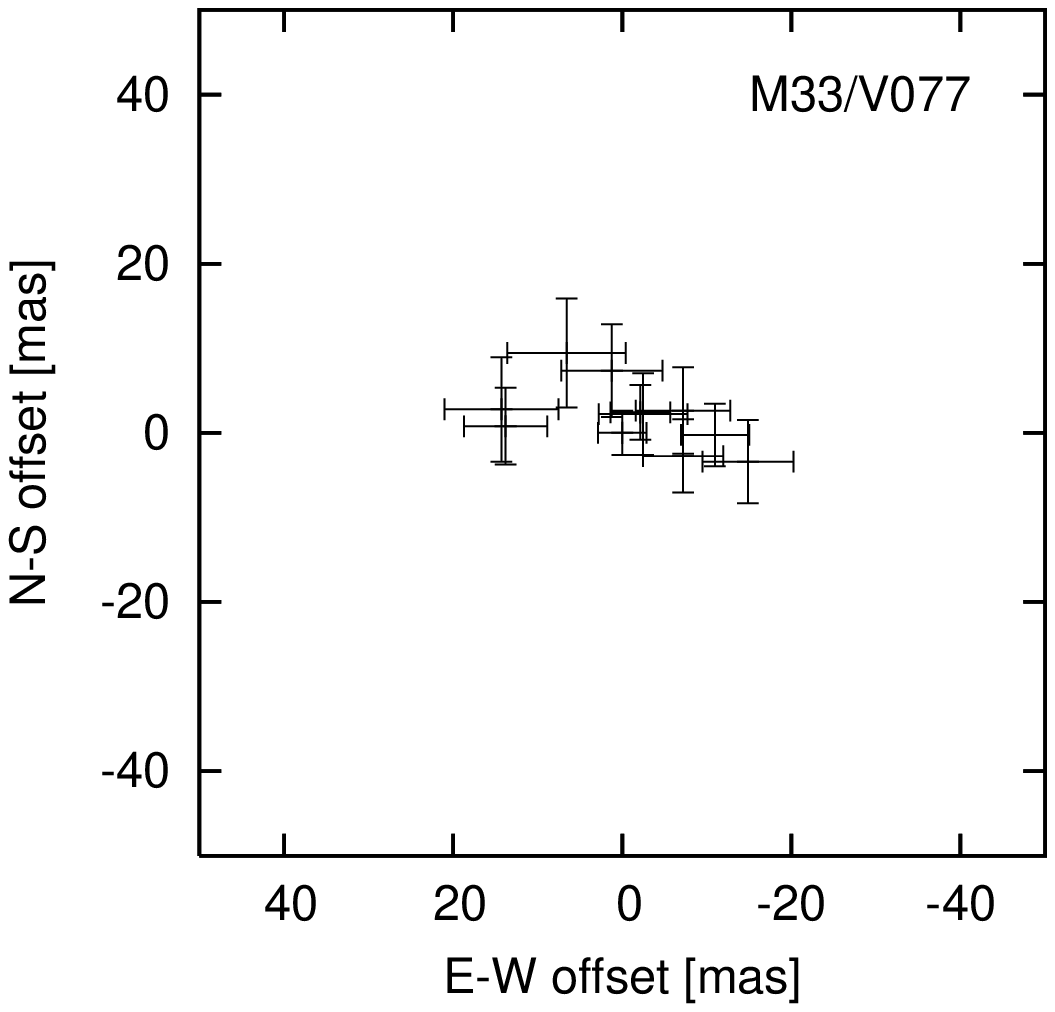}}
\caption{Positions of the detected maser features in each velocity channel relative to the position of the brightest channel from IC\,133 on 15 September 2004 (top), M33/19 on 18 November 2004 (middle), and M33/V077 on 15 September 2004 (bottom). Radial velocities for selected channels are given in brackets. The errorbars are statistical errors.}
\label{m33-pos}
\end{figure}

Fig.~\ref{m33-pos} shows the positions 
of the detected maser features in each velocity channel relative to the 
position of the brightest channel. In M33/19 and M33/V077 the emission in
all frequency channels was found less than 20 miliarcseconds from the position 
of the brightest channel. In IC\,133-Main one can clearly see that the emission
is spatially extended with an extension towards the south east. Higher 
resolution VLBI observations of IC\,133-Main have shown that IC\,133-Main 
consists of two spatially separated regions of maser activity 
\cite{GreenhillMoranReid1990,Brunthaler2004,Brunthaler2005}, one with the strong emission at $-210$ 
km~s$^{-1}$ and one is located $\sim$0.02 arcseconds towards the south-east with emission at velocities of $-220$ to $-240$ km~s$^{-1}$. The 
emission at $-171$ km~s$^{-1}$ and $-243$ km~s$^{-1}$ are located in the 
south-eastern and north-eastern tip of the region with maser emission, 
respectively.

In M33/V077 we also detect a weaker component at $\sim-188$ km~s$^{-1}$ that 
was not seen before. Follow-up VLBI observations will be needed to see 
whether or not this component originates toward the same region on scales
of a few milliarcseconds as the $-215$ km~s$^{-1}$ complex.

\subsection{NGC\,6822}

No maser was detected in NGC\,6822. We derived 5$\sigma$ upper limits 
between 36 and 78 mJy. On 31 March 2006 we detected an 8$\sigma$ peak in one 
frequency channel towards NGC\,6822-S13. However, we do not consider this a 
firm detection, since it was found in only one channel and not confirmed in the
observation of 10 April 2006. Please note that 5-7$\sigma$ peaks were found 
in almost every source, because of the large number ($\sim10^6$) of 
independent beams for each pointing, while in the case of the high velocity 
components in IC\,133 we have one additional constraint -- the position.

\section{Discussion}

\begin{figure}
\resizebox{\hsize}{!}{\includegraphics[]{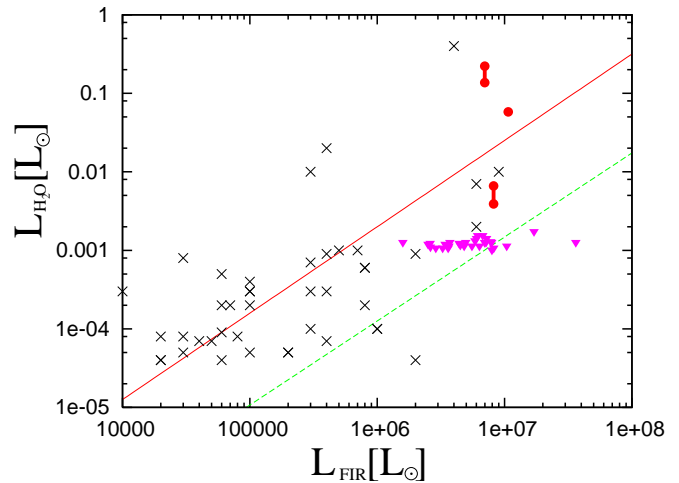}}
\caption{Maser luminosity versus FIR luminosity in M33. The FIR luminosities of
the respective H{\sc ii} regions were taken from 
\protect\citeN{HippeleinHaasTuffs2003}. The lines represent the correlations 
found for Galactic maser sources by \protect\citeN{PalagiCesaroniComoretto1993}
for detected sources (solid line) and including non detections (dashed line). 
The triangles are 5$\sigma$ upper limits. The masers in IC\,133 and M33/V077 
were observed twice. Crosses (x) are the Galactic water masers from 
\protect\citeN{PalagiCesaroniComoretto1993}.
}
\label{correl}
\end{figure}

In Fig.~\ref{correl}, we plot the maser luminosity (isotropic emission assumed) versus 
the far-infrared (FIR) luminosity for the sources in our observations of M33. 
The FIR luminosities of the H{\sc ii} regions were taken from 
\citeN{HippeleinHaasTuffs2003}. The FIR data in \citeN{HippeleinHaasTuffs2003}
has a spatial resolution of 240-400 pc in the 60-170 $\mu$m range. Also 
plotted are the correlations found for Galactic water maser sources from 
\citeN{PalagiCesaroniComoretto1993}. The solid line shows the correlation 
for detected maser sources. The dashed line shows the correlation if one 
considers also the Galactic non detections. Comparing the FIR luminosities 
of sources in M33 with Galactic sources is difficult, because of the low 
angular resolution of the FIR data from M33. 
The FIR luminosities will be overestimated if several H{\sc ii} regions are 
blended within one beam. Using higher resolution (i.e. Spitzer) data might 
therefore move the points in Fig.~\ref{correl} to the left. Hence, the FIR
luminosity values in Fig.~\ref{correl} are upper limits. Nevertheless, our 
results are in rough agreement with the brigthest masers of the Milky Way, 
given the large scatter of more than one magnitude in the Milky Way.

The detection limit for our search in M33 corresponds to 0.0015 L$_\odot$ 
(47 mJy in three channels of 0.7 km~s$^{-1}$ width) for an assumed 
distance of 809 kpc (\citeNP{McConnachieIrwinFerguson2005}). 
\citeN{GreenhillMoranReid1990} derive the luminosity function for the 
H$_2$O masers in Galactic star forming regions to be
\begin{equation}
\mathrm{log}~N=-0.6\times(1+\mathrm{log}~L_\mathrm{H_{2}O})
\end{equation}

Using this luminosity function, one expects $\sim$ 12 masers with 
$L_\mathrm{H_{2}O}>$0.0015 L$_\odot$ in the Milky Way. This is reasonable 
close to the 9 known masers with $L_\mathrm{H_{2}O}>$0.0015 L$_\odot$ in 
the Milky Way from the list in \citeN{PalagiCesaroniComoretto1993}. Repeating 
this analysis 
using the flux density detection limit instead of the maser luminosity 
detection limit yields similar but slightly lower numbers. For our Galaxy at 
the distance to M33 one would expect 8 masers with peak flux densities above
47 mJy. The slightly different value can be explained by sources with a 
total luminosity larger than 0.0015 L$_\odot$ that show broad line profiles 
but have a lower peak flux. These sources would not be detected by our 
observations. 

To better compare this with our detection rate in M33, one should account for 
the 
different star formation rates in the two galaxies. While the star formation 
rate in M33 is 0.33 -- 0.69 M$_\odot$~yr$^{-1}$ 
(\citeNP{HippeleinHaasTuffs2003}), the Milky Way has a star formation rate of 
$\sim$ 4 M$_\odot$~yr$^{-1}$ (\citeNP{DiehlHalloinKretschmer2006}) while 
other studies give values between 0.8 M$_\odot$~yr$^{-1}$ (\citeNP{Talbot1980})
and 13 M$_\odot$~yr$^{-1}$ (\citeNP{GuestenMezger1982}). Assuming that the 
maser rate and star formation rate are roughly proportional, which is 
reasonable when cloud conditions and initial stellar mass functions are 
similar, one would expect a factor of $\frac{4}{0.69}=6$ to 
$\frac{4}{0.33}=12$ times less 
masers in M33 than in the Milky Way. For our detection limit of 0.0015 
L$_\odot$ we thus expect to find $1 - 2$ masers in M33. For the flux density 
detection limit of 47 mJy one expects to find $0.7 - 1.3$ masers in M33.

Our detection limit for NGC\,6822 corresponds to 0.0008 
L$_\odot$ for a distance of 490 kpc (\citeNP{Mateo1998}) and is a factor 10 better than previous searches \cite{GreenhillMoranReid1990}. Our Galaxy hosts 
18 or 15 water masers with $L_\mathrm{H_{2}O}>$0.0008 L$_\odot$ or a peak 
flux density of more than 50 mJy, when viewed from a distance of 490 kpc. 
Scaled with the star formation rate of NGC\,6822, which is 0.06 
M$_\odot$~yr$^{-1}$ (\citeNP{deBlokWalter2006}), one expects to find 
$\frac{0.06}{4}\times 18=0.27$ masers in NGC\,6822.

\begin{table}
\begin{center}
\caption{Local Group galaxies that were searched for masers with the 
detection limit, number of detected masers, star formation rate, and number 
of expected masers.}
\label{galaxies}
\begin{tabular}{c|r|r|c|r}
Source& $L_\mathrm{H_2O}$ [L$_\odot$]&N$_\mathrm{H_2O}$&SFR [M$_\odot$~yr$^{-1}$]&N$_\mathrm{expected}$\\
\hline\hline
M33       & $1.5\times 10^{-3}$ $^\mathrm{a)}$   & 3 & 0.33-0.69 $^\mathrm{h)}$ &  1--2  \\
NGC\,6822 & $8.0\times 10^{-4}$ $^\mathrm{a)}$   & 0 & 0.06      $^\mathrm{i)}$ &  0.3   \\
IC\,10    & $4.0\times 10^{-3}$ $^\mathrm{b)}$   & 2 & 0.05      $^\mathrm{j)}$ &  0.1   \\
SMC       & $3.5\times 10^{-4}$ $^\mathrm{c)}$   & 2 & 0.05      $^\mathrm{k)}$ &  0.4   \\
LMC       & $3.0\times 10^{-5}$ $^\mathrm{d)}$   & 7 & 0.40      $^\mathrm{k)}$ & 13.0   \\
M31       & $4.9\times 10^{-3}$ $^\mathrm{e,f)}$ & 0 & 0.35      $^\mathrm{l)}$ &  0.5   \\
IC\,1613  & $9.0\times 10^{-3}$ $^\mathrm{e)}$   & 0 & $2.5\times 10^{-3}$   $^\mathrm{j)}$ &  $3\times 10^{-3}$ \\
DDO\,187  & $3.0\times 10^{-2}$ $^\mathrm{e)}$   & 0 & $3\times 10^{-4}$  $^\mathrm{j)}$ &  $1.5 \times 10^{-4}$ \\
GR8       & $2.1\times 10^{-2}$ $^\mathrm{e)}$   & 0 & $2.2\times 10^{-3}$  $^\mathrm{j)}$ &  $1.4 \times 10^{-3}$\\
NGC\,185  & $4.4\times 10^{-3}$ $^\mathrm{g)}$   & 0 & $3-6\times 10^{-3}$ $^\mathrm{m)}$ & 0.005 - 0.01\
\end{tabular}
\end{center}
References: a) this work
            b) \citeNP{BeckerHenkelWilson1993};
            c) \citeNP{ScaliseBraz1982};
            d) \citeNP{LazendicWhiteoakKlamer2002};
            e) \citeNP{GreenhillHenkelBecker1995}
            f) \citeNP{ImaiIshiharaKameya2001};
            g) \citeNP{HuchtmeierRichterWitzel1980};
            h) \citeNP{HippeleinHaasTuffs2003};
            i) \citeNP{deBlokWalter2006};
            j) \citeNP{HunterElmegreen2004};
            k) \citeNP{Hatzidimitriou1999,Israel1980};
            l) \citeNP{WalterbosBraun1994};
            m) \citeNP{Martinez-DelgadoAparicioGallart1999}
\end{table}

We can repeat this analysis for other Local Group galaxies with water maser searches from the literature, namely IC\,10, the Magellanic Clouds, M31, IC\,1613, DDO\,187, NGC\,185, and GR8. The observational detection limits, observed detection rates, estimated star formation rates, and number of expected maser sources for the above mentioned galaxies are summarized in Table~\ref{galaxies}. The searches for maser sources in any of the star forming Local Group galaxies, except DDO187 and GR8 are not complete due to the large angular size of the galaxies. Thus, it is possible that additional masers are located in regions which were not measured. However, in the Milky Way the most luminous water masers are found in the 
largest star forming regions (see e.g. the maser luminosity - FIR luminosity correlation in \citeNP{PalagiCesaroniComoretto1993}). Since the largest star forming regions were targeted in the maser searches at least in M33, M31, IC\,1613, and NGC\,6822 it is not likely that additional masers exist above the detection limits mentioned in Table~\ref{galaxies}.

The number of masers in all galaxies except IC\,10 is consistent with the expected number of maser sources. However, IC\,10 shows an overabundance. Possibly, star formation in this galaxy is concentrated in two particular regions near the center of the galaxy, where 
a small 'starburst' creates conditions different from those expected from 
an average galaxy containing little molecular gas. Based on this analysis it is also not surprising that the previous maser searches in M31, IC\,1613, DDO\,187, and GR8 with flux density limits of a few hundred mJy, did not detect any masers.


\section{Conclusion}

We have searched for water maser emission with the VLA toward 62 and 36 
H{\sc ii} regions in M33 and NGC\,6822, respectively. M33 hosts three water 
masers above our detection limit (all previously identified), while in NGC\,6822 no maser source was 
detected. IC\,133 and M33/V077 show velocity components that are displaced 
by several 10 km~s$^{-1}$ from previously observed maser features. We also 
provide an accurate position for M33/V077, while several previously reported 
marginal detections in M33 are not confirmed.

The number of known maser sources in the Local Group galaxies M31, M33, 
NGC\,6822, SMC, LMC, IC\,1613, DDO\,187, NGC\,185, and GR8 is consistent with 
the expected number of maser sources based on the Galactic water maser 
luminosity function and the current star formation rates. Only the galaxy 
IC\,10 shows a much higher detection rate than expected.

Based on this analysis one can expect to find 17 and 5 water masers with 
flux densities above 1 mJy in M31 and NGC\,6822, respectively, that could 
be used to measure the proper motions and geometric distances of these 
galaxies with future radio telescopes like the Square Kilometre Array 
(SKA).\footnotemark[2]

\footnotetext[2]{See http://www.skatelescope.org.}

\appendix{}
\section{Tables}

\begin{table*}
\begin{center}
\caption{List of observed H{\sc ii} regions on 14 September 2004 with observed coordinates,
LSR velocity of the band center and the peak flux density or the 5$\sigma$ limit for a single channel.
\label{source1}}
\begin{tabular}{rccccr }
\hline\hline
 Source & Other & RA  & DEC & v$_{LSR}$ & F$_{peak}$\\   
        & & (B1950) & (B1950)    & [km s$^{-1}$]  &  [mJy]     \\
\hline
M33/V067 & M33/42$^1$ & 01 31 03.50 & +30 23 55.0 & -174& $<$ 43\\
M33/V068 & & 01 31 03.60 & +30 28 26.0 &-248& $<$ 42 \\
M33/V069 & & 01 31 07.60 & +30 17 30.0 &-90 & $<$ 52 \\
M33/V070 & IC\,142 & 01 31 05.80 & +30 30 00.0 &-244 & $<$ 43\\
M33/V072 & M33/46$^2$ & 01 31 08.60 & +30 26 50.0 &-252 & $<$ 43\\
M33/V075 & & 01 31 10.10 & +30 20 29.0 &-151 & $<$ 40\\
M33/V076 & & 01 31 10.20 & +30 19 03.0 &-140 & $<$ 42\\
M33/V077 & M33/50\footnotemark[1] & 01 31 10.90 & +30 25 29.0 &-209 & 111\\
M33/V028 & IC\,133 &  01 30 27.20 & +30 37 30.0 &-210 & 2150\\
M33/V079 & M33/51$^1$ & 01 31 13.00 & +30 23 19.0 &-198 & $<$ 37\\
M33/V083 & & 01 31 17.30 & +30 26 27.0 &-232 & $<$ 40\\
M33/V084 & & 01 31 17.40 & +30 33 33.0 &-278 & $<$ 37\\
M33/V085 & & 01 31 15.30 & +30 31 48.5 &-271 & $<$ 38\\
M33/V089 & & 01 31 21.00 & +30 23 41.0 &-198 & $<$ 38\\
M33/V091 & & 01 31 21.50 & +30 27 04.0 &-181 & $<$ 43\\
M33/V092 & & 01 31 21.60 & +30 21 01.0 &-164 & $<$ 42\\
M33/V094 & & 01 31 24.90 & +30 19 13.0 &-174 & $<$ 53\\
M33/V095 & M33/58$^1$ & 01 31 26.50 & +30 21 49.0 &-212 & $<$ 48\\
M33/V099 & M33/61$^1$ & 01 31 28.10 & +30 18 27.0 &-160 & $<$ 46\\
M33/V103 & & 01 31 33.50 & +30 17 59.0 &-167 & $<$ 42\\
M33/V106 & NGC\,604  & 01 31 43.60 & +30 31 44.0 &-264 & $<$ 43\\
M33/V109 & & 01 31 49.60 & +30 29 00.0 &-225 & $<$ 39\\
M33/V110 & M33/67 & 01 31 50.30 & +30 26 19.0 &-197 & $<$ 40\\
M33/V111 & & 01 31 51.70 & +30 32 43.0 &-232 & $<$ 40\\
M33/V112 & & 01 31 51.80 & +30 28 01.0 &-148 & $<$ 41\\
M33/V005 & & 01 29 53.30 & +30 05 41.0 &-105 \footnotemark[2] & $<$ 41\\
M33/V046 & & 01 30 47.60 & +30 05 04.0 &-104 \footnotemark[2] & $<$ 42\\
M33/V052 & & 01 30 52.50 & +30 05 36.0 &-106 \footnotemark[2] & $<$ 45\\
M33/V097 & M33/60$^1$ & 01 31 27.10 & +30 36 55.0 &-254 \footnotemark[2] & $<$ 40\\
M33/V104 & & 01 31 40.20 & +30 41 39.0 &-259 \footnotemark[2] & $<$ 39\\
M33/V107 & M33/65$^1$ & 01 31 44.10 & +30 45 00.0 &-263 \footnotemark[2] & $<$ 40 \\
\hline
\end{tabular}

$^1$ After designation of \citeN{IsraelvanderKruit1974}\\
$^2$ using the rotation model of \citeN{CorbelliSchneider1997}
\end{center}
\end{table*}

\begin{table*}
\begin{center}
\caption{List of observed H{\sc ii} regions on 18 November 2004 with observed coordinates,
LSR velocity of the band center and the peak flux density or the 5$\sigma$ limit for a single channel.\label{source2}}
\begin{tabular}{rccccr }
\hline\hline
 Source & Other & RA  & DEC & v$_{LSR}$ & F$_{peak}$\\   
        & &  (B1950)   & (B1950)    & [km s$^{-1}$] &   [mJy]    \\
\hline
M33/V001 & & 01 29 41.50 & +30 12 17.0 &-147 & $<$ 48\\
M33/V003 & & 01 29 45.00 & +30 12 07.0 &-135 & $<$ 47\\
M33/V006 & & 01 29 55.70 & +30 19 36.0 &-158 & $<$ 48\\
M33/V007 & NGC\,588 & 01 29 56.50 & +30 23 30.0&-158 \footnotemark[2]& $<$ 47\\
M33/V013 & M33/2\footnotemark[1] & 01 30 07.70 & +30 23 30.0 &-135 & $<$ 48\\
M33/V019 & & 01 30 20.40 & +30 07 51.0 &-90 & $<$ 45\\
M33/V020 & M33/5$^1$ & 01 30 21.20 & +30 12 07.0 &-107 & $<$ 48\\
M33/V021 & IC\,131  & 01 30 22.20 & +30 29 53.0 &-212 & $<$ 46\\
M33/V022 & NGC\,592 & 01 30 22.90 & +30 23 19.0 &-168 & $<$ 46\\
M33/V023 & & 01 30 23.00 & +30 14 50.0 &-103 & $<$ 56\\
M33/V026 & & 01 30 25.40 & +30 29 49.0 &-200 & $<$ 58\\
M33/V027 & IC\,132 & 01 30 26.70 & +30 41 21.0 &-242 & $<$ 52\\
M33/V028 & IC\,133 & 01 30 27.20 & +30 37 30.0 &-210 & 1600\\
M33/V029 & & 01 30 29.80 & +30 26 08.0 &-191 & $<$ 57 \\
M33/V033 & M33/19\footnotemark[1] & 01 30 39.20 & +30 16 21.0 &-135 & 459\\
M33/V034 & M33/18$^1$ & 01 30 39.80 & +30 25 00.0 &-180 & $<$ 58\\
M33/V037 & & 01 30 42.30 & +30 18 13.0 &-139 & $<$ 62\\
M33/V040 & M33/21$^1$& 01 30 44.50 & +30 16 46.0 &-119 & $<$ 65\\
M33/V041 & NGC\,595 & 01 30 44.80 & +30 26 08.0 &-220 &$<$ 65 \\
M33/V042 & M33/23$^1$ & 01 30 46.50 & +30 27 07.0 &-170 & $<$ 58\\
M33/V043 & & 01 30 46.60 & +30 24 18.0 &-140 & $<$ 48\\
M33/V044 & M33/25$^1$ & 01 30 46.70 & +30 21 07.0 &-127 & $<$ 53\\
M33/V045 & & 01 30 47.30 & +30 35 18.0 &-241 & $<$ 51\\
M33/V049 & & 01 30 50.30 & +30 22 41.0 &-131 & $<$ 49\\
M33/V050 & & 01 30 51.10 & +30 30 37.0 &-222 & $<$ 44\\
M33/V054 & & 01 30 53.30 & +30 17 50.0 &-136 & $<$ 48\\
M33/V057 & M33/30$^1$ & 01 30 55.20 & +30 29 22.0 &-222 & $<$ 51\\
M33/V059 & & 01 30 56.10 & +30 21 23.0 &-135 & $<$ 51\\
M33/V060 & & 01 30 58.30 & +30 28 24.0 &-221 & $<$ 49\\
M33/V061 & M33/35$^1$ & 01 30 58.40 & +30 23 20.0 &-171 & $<$ 44 \\
M33/V063 & M33/38$^1$ & 01 30 59.40 & +30 24 17.0 &-212 & $<$ 44\\
M33/V064 & M33/39$^1$ & 01 31 00.60 & +30 22 07.0 &-130 & $<$ 44\\
M33/V077 & M33/50\footnotemark[1] & 01 31 11.00 & +30 25 23.0 &-215 & 83\\
\hline
\end{tabular}

$^1$ After designation of \citeN{IsraelvanderKruit1974}\\
$^2$ using the rotation model of \citeN{CorbelliSchneider1997}
\end{center}
\end{table*}

\begin{table*}
\begin{center}
\caption{List of observed H{\sc ii} regions on 18 November 2004 with observed coordinates,
LSR velocity of the band center and the  5$\sigma$ limit for a single channel.\label{ngc6822_source}}
\begin{tabular}{rrccccr }
\hline\hline
 Date& Source & Other & RA  & DEC & v$_{LSR}$ & F$_{peak}$\\   
       & & &  (J2000)   & (J2000)    & [km s$^{-1}$] &   [mJy]    \\
\hline
13.03.2006 & NGC\,6822-01 && 19:45:05.04974&-14:43:12.6178&-48&$<$43\\
           & NGC\,6822-02 && 19:44:52.86152&-14:43:11.9484&-58&$<$44\\
           & NGC\,6822-03 && 19:44:48.13213&-14:44:17.8702&-68&$<$45\\
           & NGC\,6822-04 && 19:45:17.10323&-14:45:29.4447&-38&$<$46\\
           & NGC\,6822-05 && 19:44:34.27118&-14:42:16.2557&-78&$<$47\\
           & NGC\,6822-06 && 19:44:31.51770&-14:41:56.9218&-78&$<$49\\
           & NGC\,6822-07 && 19:44:31.87935&-14:44:31.4359&-78&$<$51\\
\hline
25.03.2006 & NGC\,6822-08 && 19:44:48.54365&-14:46:12.0904&-65&$<$36\\
           & NGC\,6822-09 && 19:44:32.83735&-14:47:27.9327&-75&$<$38\\
           & NGC\,6822-10 && 19:44:30.90656&-14:48:23.8619&-75&$<$39\\
           & NGC\,6822-11 && 19:45:13.06248&-14:45:14.1431&-40&$<$40\\
           & NGC\,6822-12 && 19:44:38.20420&-14:51:01.7538&-65&$<$38\\
           & NGC\,6822-16 && 19:45:12.65272&-14:48:47.9304&-40&$<$38\\
           & NGC\,6822-17 && 19:45:06.86589&-14:48:35.9576&-40&$<$41\\
\hline
31.03.2006 & NGC\,6822-13 && 19:45:01.90566&-14:46:52.0675&-55&$<$48\\
           & NGC\,6822-14 && 19:44:56.94607&-14:48:01.9967&-55&$<$49\\
           & NGC\,6822-15 && 19:44:52.53709&-14:48:33.9585&-55&$<$49\\
           & NGC\,6822-18 && 19:44:55.84312&-14:50:29.8476&-55&$<$51\\
           & NGC\,6822-19 && 19:44:50.32947&-14:52:51.6951&-55&$<$49\\
           & NGC\,6822-20 && 19:44:52.14471&-14:51:54.7565&-55&$<$59\\
           & NGC\,6822-23 && 19:45:09.77979&-14:45:25.8082&-40&$<$63\\
           & NGC\,6822-24 && 19:44:48.84296&-14:45:12.4843&-65&$<$59\\
\hline
10.04.2006 & NGC\,6822-21 && 19:45:04.80200&-14:57:13.1100& -40&$<$58\\
           & NGC\,6822-22 && 19:45:13.62800&-14:58:59.6500& -40&$<$58\\
           & NGC\,6822-25 && 19:45:01.49300&-14:54:27.9400& -40&$<$58\\
           & NGC\,6822-28 && 19:45:11.05000&-14:54:38.5800& -40&$<$58\\
           & NGC\,6822-29 && 19:45:12.70700&-14:57:26.4100& -40&$<$60\\
           & NGC\,6822-13 && 19:45:02.48330&-14:47:09.6170& -70&$<$46\\
\hline
17.04.2006 & NGC\,6822-26 && 19:44:48.99900&-14:50:33.5000& -60&$<$45\\
           & NGC\,6822-27 && 19:44:42.56900&-14:50:04.1700& -60&$<$45\\
           & NGC\,6822-30 && 19:44:58.80600&-14:43:57.2400& -50&$<$37\\
           & NGC\,6822-31 && 19:45:02.84700&-14:45:14.5000& -50&$<$42\\
           & NGC\,6822-32 && 19:44:56.23500&-14:45:43.8000& -50&$<$38\\
           & NGC\,6822-33 && 19:45:12.76200&-14:43:33.2500& -50&$<$43\\
           & NGC\,6822-34 && 19:44:56.23600&-14:42:08.0200& -50&$<$43\\ 
           & NGC\,6822-35 && 19:44:15.64400&-14:46:15.5100& -80&$<$46\\
           & NGC\,6822-36 && 19:44:39.34200&-14:43:06.5700& -80&$<$48\\
\hline
\end{tabular}
\end{center}
\end{table*}

\begin{acknowledgements}
This research was supported by the DFG Priority Programme 1177.
\end{acknowledgements}

\bibliography{brunthal_refs}
\bibliographystyle{aa}

\end{document}